\RequirePackage{fix-cm}
\documentclass[smallextended]{svjour3}       
\smartqed  
\usepackage{graphicx}
\usepackage{amsmath}
\newcommand{\ea}{{\em et al.}}
\newcommand{\Cs}{{\ensuremath{\mathcal{C}}}}
\newcommand{\Ps}{{\ensuremath{\mathcal{P}}}}
\newcommand{\Ts}{{\ensuremath{\mathcal{T}}}}
\newcommand{\ecm}{\ensuremath{e\cdot\text{cm}}}
\journalname{Hyperfine Interactions}
\begin{document}

\title{Search for Electric Dipole Moments at Storage Rings }

\author{C.J.G.  Onderwater }

\institute{Gerco Onderwater \at
              KVI, University of Groningen \\
              Zernikelaan 25, NL-9747AA Groningen, the Netherlands\\
              \email{onderwater@kvi.nl}           
}

\date{Received: date / Accepted: date}

\maketitle

\begin{abstract} Permanent electric dipole moments (EDMs) violate
parity and time-reversal symmetry.  Within the Standard Model (SM)
they are many orders of magnitude below present experimental
sensitivity.  Many extensions of the SM predict much larger EDMs,
which are therefore an excellent probe for the existence of ``new
physics''.  Until recently it was believed that only electrically
neutral systems could be used for sensitive searches of EDMs.  With
the introduction of a novel experimental method, high precision for
charged systems will be within reach as well.  The features of this
method and its possibilities are discussed.   \end{abstract}

\section{Motivation}

The symmetry properties of fundamental processes and particles are a
strong guide to understand the underlying interactions.  The QED
Lagrangian in the current Standard Model (SM) predicts that all
electromagnetic observables are even under the discrete symmetries \Cs
(charge conjugation), \Ps\ (parity) and \Ts\ (time reversal)
individually and thus under each combinations of them.  Strong
interaction observables, described by QCD, are also predicted to be
even under \Cs, \Ps\ and \Ts, with the exception of those proportional
to $\bar{\theta}$ which are \Ps\ and \Ts-odd\cite{Cheng1988}.  The
weak interaction Lagrangian is odd under \Ps\ and \Cs\ because of the
handedness of the coupling of the W and Z-bosons.  Many observables
are even under their combination \Cs\Ps.  Nevertheless, the weak
interaction also predicts \Cs\Ps-odd ones.  These are all proportional
to the Jarlskog invariant, $J \propto
\sin^2\theta_{12}\sin\theta_{23}\sin\theta_{13}\sin\delta \sim 3\times
10^{-5}$\cite{Jarlskog:1985cw}.  Here $\theta_{12}$, $\theta_{23}$ and
$\theta_{13}$ are the quark flavor mixing angles and $\delta$ the
\Cs\Ps-violating complex phase associated with the Kobayashi-Maskawa
mechanism\cite{Kobayashi:1973fv}.  The SM is built on the assumption
of Lorentz invariance and hence the invariance of the combination
\Cs\Ps\Ts.  Consequently \Ts\ and \Cs\Ps\ violation are equivalent.

In the SM there are thus two sources of \Cs\Ps/\Ts\ violation,
$\bar{\theta}_{QCD}$ and $J$.  The magnitude of the former is
as-of-yet undetermined, $|\bar{\theta}|<\mathcal{O}(10^{-11})$,
whereas the smallness of the latter guarantees that SM \Cs\Ps-odd
observables are generally small.

Violation of \Cs\Ps\ is also expected to be necessary to explain hte
baryon asymmetry (BAU) in the universe\cite{Sakharov:1967}.  The BAU
predicted from the SM and the Cosmological Standard Model falls short
of the observed one by as much as ten order of magnitude.  This
suggests the presence of additional sources of \Cs\Ps-violation beyond
those incorporated in the SM.

Permanent electric dipole moments (EDMs) are an excellent tool to
search for such addition sources of \Cs\Ps-violation\cite{nmtest}. 
EDMs break both \Ps\ and \Ts, which is manifest when considering the
field dependent part of the interaction Hamiltonian $\mathcal{H}$ for
a particle in an electric field $\vec{E}$ and a magnetic field
$\vec{B}$,
\begin{equation}
  \mathcal{H}=  -\left(\vec{\mu}\cdot\vec{B} + \vec{d}\cdot\vec{E} \right)
      = -\left(\mu\vec{B} + d\vec{E} \right)\cdot\frac{\vec{J}}{J}.
\end{equation}
The second equality holds because the spin $\vec{J}$ is the only
vector in the rest frame of a fundamental particle.  The electric and
magnetic dipole moments must point along it; $\mu=g(e\hbar/2m)$ and
$d=\eta(e\hbar/4mc)$ are the respective proportionality constants. 
$g$ and $\eta$ are the respective dimensionless moments and
$e\hbar/2m$ the Bohr magneton.  When $d\not=0$ this Hamiltonian
violates both \Ps\ and \Ts.

Particles acquire non-zero EDMs through \Cs\Ps\ violating radiative
corrections.  Quark mixing, {\em i.e.} $\delta$, contributes only
through three or more weak-interaction loop corrections.  This makes
these EDMs extremely small, of order $10^{-31}$\,\ecm\ for hadronic
systems down to $10^{-41}$\,\ecm\ for leptons.  This is far below
present detection limits\cite{Khriplovich:1997}.  Hadronic systems may also
acquire an EDM through $\bar{\theta}$ without the need for multiple
loops.  The non-observation of the neutron EDM limits the magnitude of
$|\bar{\theta}| < \mathcal{O}(10^{-11})$.

In many proposed extensions of the SM, the need for multiple loops is
not present, and EDMs may occur even at first
order\cite{Sandars:2001}.  For example, many supersymmetric (SUSY)
models predict a neutron EDM
\begin{equation}
  d_n(SUSY) \sim \sin\delta_{CP} \left(\frac{1\,\text{TeV}}{M_{SUSY}} \right)^2 \times 10^{-25}\,\ecm.
\end{equation}
Once the masses of the SUSY particle are determined, the \Cs\Ps\
violating phase $\delta_{CP}$ can be determined using EDMs.

The first observation of a non-zero EDM would already unambiguously
establish the presence of physics beyond the SM.  Different forms of
new physics manifest themselves differently already at the level of
the fundamental fermions and leptons, and their interactions, and
propagate into increasingly larger composite systems, from hadrons to
nuclei, atoms and molecules\cite{hierarchy}.  At each stage the
appropriate theory needs to be applied.  The most stringent limits on
quark and proton EDM, as well as on \Cs\Ps-violating electron-nucleus
interactions are derived from the EDM search on the $^{199}$Hg
atom\cite{Griffith:2009zz}.  The most strict electron EDM limit stems
from the YbF molecule\cite{Hudson:2011zz}.  The muon EDM is the only
fundamental particle for which the EDM was obtained
directly\cite{Bennett:2008dy}.

A single EDM measurement cannot be traced back to a specific source of
\Cs\Ps-violation.  Several complementary EDM measurements are
necessary.  For example, the combination of the neutron EDM combined
with that for the proton, deuteron and possibly helion and triton
makes it possible to distinguish new sources of \Cs\Ps-violation from
that introduced by $\bar{\theta}$\cite{DeVries2011b}.  The
uncertainties in the theory to describe light nuclei are well under
control permitting reliable
predictions\cite{TimmPC,Afnan2010,DeVries2011}.  Light nuclei thus
offer the theoretically cleanest way to study hadronic \Cs\Ps\
violation.

Light nuclei cannot be probed for an EDM using atom- or molecule-based
methods because of shielding effects\cite{Schiff}.  Several
experimentals methods that circumvent these problems make use of the
motional electric field a fast moving particle experiences when
traversing a magnetic
field\cite{Berley:1960,Bailey:1977sw,Farley:2003wt,Orlov:2006}.  These
methods provide direct access to the very interesting realm of light
nuclei, which so far have not been examined for EDMs.  Also the muon
can be probed sensitively, offering the unique possibility to explore
the flavor structure of fundamental particles.

\section{Storage Ring Techniques}

A charged particle with magnetic moment anomaly $a$ and EDM $\eta$
moving in a electromagnetic field will exhibit spin precession.  The
evolution of the spin $\vec{S}$ is described by the (simplified) BMT
equation\cite{BMT},
\begin{equation}
  \cfrac{d\vec{S}}{dt} = \cfrac{e}{m}\,\vec{S}\times
    \left[ a\vec{B} +
      \left(\frac{1}{\gamma^2-1}-a\right)\vec{\beta}\times\vec{E} + 
      \cfrac{\eta}{2}\,\left(\vec{E} + \vec{\beta}\times\vec{B}\right)
    \right] \equiv \vec{S}\times\vec{\Omega}.
  \label{eq:gm2andEDM}
\end{equation}
The first two terms arrise from the interaction of the magnetic moment
with the magnetic field, whereas the last term is due to the
interaction of the EDM with the electric field.  The EDM interacts
with the electric field in the rest-frame of the particle, which may
have a strength $E^{CM}\sim E+ vB \sim$~GV/m far in excess of those
attainable in the laboratory.

In a purely magnetic storage ring the spin precesses about
$\vec{\Omega}$ which is tilted with respect to $\vec{B}$ by an angle
$\psi \simeq \eta\beta/2a$.  The precession rate increases to
$\Omega = \sqrt{1 + \psi^2}\Omega_0$ with $\Omega_0 = a(e/m)B$. 
This quadratic sensitivity of the precession rate precludes a
sensitive measurement of $\eta$.  Because $\vec{\Omega}$ is tilted
with respect to $\vec{B}$ the spin component along the magnetic field
oscillates with an amplitude that depends linearly on the EDM.  In the
muon g-2 experiment at BNL this was used to limit the muon
EDM\cite{Bennett:2008dy}.  It will be used again in the new muon g-2
experiment at Fermilab\cite{Carey:2009zzb}.  The muon EDM limit of
$10^{-19}$\,\ecm\ corresponds to a tilt in the precession plane of
order 1\,microradian.

The statistical power of an EDM experiment is proportional to
${PE\sqrt{N}TA}$ with polarization $P$, effective electric field $E$,
number of particle $N$, characteristic time scale $T$ and analyzing
power $A$.  The sensitivity of the $\vec{B}$-only method is limited by
the short precession cycle, which determines $T$.

Reducing the precession rate will prolong $T$.  This can be done by
applying a suitably chosen combination of radially oriented electric
and vertically oriented magnetic fields.  For
\begin{equation}
  \frac{E}{B} = \frac{a\beta}{1-(1+a)\beta^2}.
  \label{eq:freezing}
\end{equation}
the first two terms in eq.\,(\ref{eq:gm2andEDM}) cancel, so that
$\vec{\Omega} = \eta/2 (\vec{E}+\vec{\beta}\times\vec{B})$.  The spin
precession rate is now entirely determined by the EDM.  The spin
precesses about the electric field in the rest frame of the particle,
which is oriented radially in the laboratory frame.  The signature of
an EDM is the change of the polarization component out of the particle
orbit plane, generally along the magnetic field,

For particles with positive $a$ a ``magic'' momentum $p_{magic} =
\frac{m}{\sqrt{a}}$ exists.  At this momentum the particles can be
stored in an all-electric setup with $B=0$.  The bending radius $\rho$
of a particle with mass $m$ moving in an electric field $E$ is
\begin{equation}
  \rho = \frac{1}{\sqrt{a(a+1)}} \frac{m}{E}.
\end{equation}
For a proton $p_{magic} \simeq 700\,$MeV/$c$ and $\rho\simeq 42$\,m,
assuming $E=10$\,MV/m.  The spin precesses at a rate of
%
$
  \Omega/d = {2 E}/{\hbar} \simeq\, 10^{20}\,{\text{rad/s}}/({\ecm}).
$
%
At the expected sensitivity for $\Omega$ of 1\,nrad/s a proton EDM of
$d_{p} = 10^{-29}$\,\ecm\ can be measured.  Efforts are ongoing to
realize such an experiment\cite{Semertzidis2011}.

An all-electric setup is not feasible for particles with small $a$ and
impossible for those with a negative one.  A combination of electric
and magnetic fields is necessary to ``freeze'' the spin.  Expressed in
$E$ the EDM induced spin precession rate and corresponding bending
radius are given by
\begin{equation}
  \Omega = \frac{2 d (E + vB)}{\hbar} = \frac{a+1}{a\gamma^2} \frac{2dE}{\hbar}
  ~~~~~\text{and}~~~~~
  \rho = \frac{a\beta^2\gamma^3}{a+1} \frac{m}{E}
\end{equation}
Both from the point of view of the spin precession rate and the size
of the setup low momenta are preferred.  The electric field is
effectively amplified by $(a+1)/a\gamma^2$.  This becomes sizable for
particles with a small $a$ (see \cite{Khriplovich2000}).  For the
electron or muon with $a\simeq 0.00116$ this is $(a+1)/a \simeq 860$.

For muons with $p = 500$\,MeV/$c$ and $E=2.2$\,MV/m as proposed in
\cite{Aoki2003} $\rho= 7$\,m.  A considerably smaller setup with
$\rho=0.42$\,m is proposed in \cite{Adelmann2010} with $p =
125$\,MeV/$c$ and $E=0.64$\,MV/m.  The projected sensitivities are
about the same at $d_\mu \simeq 10^{-16}\,\ecm/\sqrt{N}$, with $N$ the
number of detected muon decays.  At exisiting muon facilities $N =
10^{12}$ could be collected yielding a statistics limited sensitivity
of $d_\mu \simeq 10^{-22}$\,\ecm.  This improves the current limit by
three orders of magnitude.  This experiment could well serve as a
small-scale low-cost demonstration of this novel technique.  At a
future high-intensity muon facility this can be further improved by
several orders of magnitude. 

At the Forschungszentrum J\"ulich the possibilities for a light-ion
EDM facility are explored\cite{COSY}.  Several options for an
``all-in-one'' storage ring were presented to search for EDMs on
protons, deuterons and $^3$He\cite{Lehrach2011}.  For a
$p_{p}=435$\,MeV/$c$, $p_{d}=1000$\,MeV/$c$ and $p_{^3\text{He}}=
765$\,MeV/$c$ a single ring with a bending radius of 10\,m can be
constructed requiring $B<0.5$\,T and $E<17$\,MV/m. 

\section{Outlook}

A sensitive EDM search using a storage ring requires besides
statistical also systematic precision.  Current R\&D efforts address
several aspects that affect both.  A system is being developped to
reliably generate the electric field strengths of 10\,MV/m planned for
the proton EDM search.  Such systems have been employed on a much
smaller scale as electrostatic separators at {\em e.g.} the AGS at BNL
and the Tevatron at FNAL.  A critical aspect is the alignment of the
electric field with respect to the magnetic field.  Spin and beam
dynamics must be understood at an unprecedent level of precision to
guarantee optimal statistican precision through a long spin coherence
time and to exclude or reduce systematic uncertainties to an
acceptable level.  An active research program is underway using the
COSY facility at the Forschungszentrum J\"ulich.  A lower limit on the
spin coherence time of 75\,s was demonstrated already, just one order
of magnitude below the goal for the proton and deuteron EDM
experiments\cite{Stephenson2011}.  Also at COSY a scheme to
efficiently measure deuteron polarization and to correct for
systematic errors was demonstrated\cite{Brantjes2012}.  The
demonstrated sensitivity is sufficient to reach the proposed
sensitivity of $d_d = 10^{-29}$\,\ecm.

In conclusion storage rings make it possible to enter new territory in
the search for EDMs.  It is expected that an experiment on the proton
and deuteron can be realized in the near future.  A first small-scale
implementation of a storage ring could be realised already now to
search for a muon EDM.  They make it possible to directly probe
charged particles with competitive sensitivity.  Such systems have a
complementary sensitivity to new sources of CP-violation and may help
to pin-down the last unconfirmed source of CP-violation in the
Standard Model, $\bar{\theta}$.

\begin{acknowledgements}
The author is indebted to his colleagues at KVI, PSI, RCNP and the
ring-EDM collaboration for valuable discussions.  The research
described in this paper was partly financed by an Innovative Research
grant (\#680-47-203) from the Dutch Organization for Fundamental
Research (NWO).
\end{acknowledgements}

\end{document}